\documentclass[10pt]{LTJournalArticle_mod}

\usepackage{amsmath}
\usepackage{amssymb}
\usepackage{enumitem}
\usepackage[utf8]{inputenc}
\usepackage{svg}

\usepackage{dblfloatfix}
\usepackage{float}
\usepackage{caption}
\usepackage{subcaption}
\usepackage{comment}
\usepackage{etoolbox}

\usepackage{xcolor}
\usepackage{graphicx}

\allowdisplaybreaks
\newcommand{\info}[1]{\textcolor{orange}{\textbf{#1}}}

\newtoggle{arxiv}
\toggletrue{arxiv}

\iftoggle{arxiv}{
\title{\LARGE RudolfV: A Foundation Model by Pathologists for Pathologists}
}{
\title{\LARGE RudolfV:\\Human-guided Pathology  Foundation Model Design}
}

\author{Jonas Dippel$^{\;*\;1,2,3}$, Barbara Feulner$^{\;*\;1}$, Tobias Winterhoff$^{\;*\;1}$, Timo Milbich$^{\;*1}$,\\ Stephan Tietz$^1$, Simon Schallenberg$^8$, Gabriel Dernbach$^{1,3,8}$, Andreas Kunft$^1$,\\ Simon Heinke$^1$, Marie-Lisa Eich$^{1,8}$,  Julika Ribbat-Idel$^1$, Rosemarie Krupar$^1$,\\ Philipp Anders$^{8}$,  Niklas Prenißl$^8$, Philipp Jurmeister$^{6, 7}$, David Horst$^{6, 8}$, Lukas Ruff$^1$,\\ Klaus-Robert Müller$^{\;\dag\;2,3,4,5}$, Frederick Klauschen$^{\;\dag\;3,6,7,8,9}$,  Maximilian Alber$^{\;\dag\;1,8}$}

\date{
    $^1$ Aignostics, Germany\\
    $^2$ Machine Learning Group,
    Technische Universität Berlin, Germany\\
    $^3$ BIFOLD – Berlin Institute for the Foundations of Learning and Data, Germany\\
    $^4$ Department of Artificial Intelligence, Korea University, Republic of Korea\\
    $^5$ Max-Planck Institute for Informatics, Germany\\
    $^6$ German Cancer Research Center (DKFZ) \& German Cancer Consortium (DKTK), Berlin \& Munich Partner Sites \\
    $^7$ Institute of Pathology, Ludwig-Maximilians-Universität München, Germany\\
    $^8$ Institute of Pathology, Charité – Universitätsmedizin Berlin, Germany\\
    $^9$ Bavarian Cancer Research Center (BZKF), Germany\\
    $^{* \dag}$ Equal Contribution
}

\renewcommand{\maketitlehookd}{%
\begin{abstract}
\normalfont
Artificial intelligence has started to transform histopathology impacting clinical diagnostics and biomedical research. However, while many computational pathology approaches have been proposed,
most current AI models are limited with respect to generalization, application variety, and handling rare diseases.
Recent efforts introduced self-supervised foundation models to address these challenges, yet existing approaches do not leverage pathologist knowledge by design.
In this study, we present a novel approach to designing foundation models for computational pathology, incorporating pathologist expertise, semi-automated data curation, and a diverse dataset from over 15 laboratories, including 58 tissue types, and encompassing 129 different histochemical and immunohistochemical staining modalities.
We demonstrate that our model ``RudolfV'' surpasses existing state-of-the-art foundation models across different benchmarks focused on tumor microenvironment profiling, biomarker evaluation, and reference case search while exhibiting favorable robustness properties.
Our study shows how domain-specific knowledge can increase the efficiency and performance of pathology foundation models and enable novel application areas.
\end{abstract}
}

\begin{document}
\clearpage
\newpage

\maketitle

\begin{comment}
\info{
Journal restrictions:
\begin{itemize}
    \item
    Main text – up to 4,000 words, excluding abstract, Methods, references and figure legends.
    \item
Abstract – up to 150 words, unreferenced.
 \item
Display items – up to 6 items (figures and/or tables).
 \item
Article should be divided as follows: Introduction (without heading), Results, Discussion, Online Methods.
 \item
Results and online Methods should be divided by topical subheadings; the Discussion does not contain subheadings.
\item
References –  as a guideline, we typically recommend up to 60.
\end{itemize}
}
\end{comment}

\section{Introduction}

%trim={2cm 1cm 2cm 4cm}
\begin{figure*}[h!]
    \centering
    \includegraphics[width=.95\textwidth]{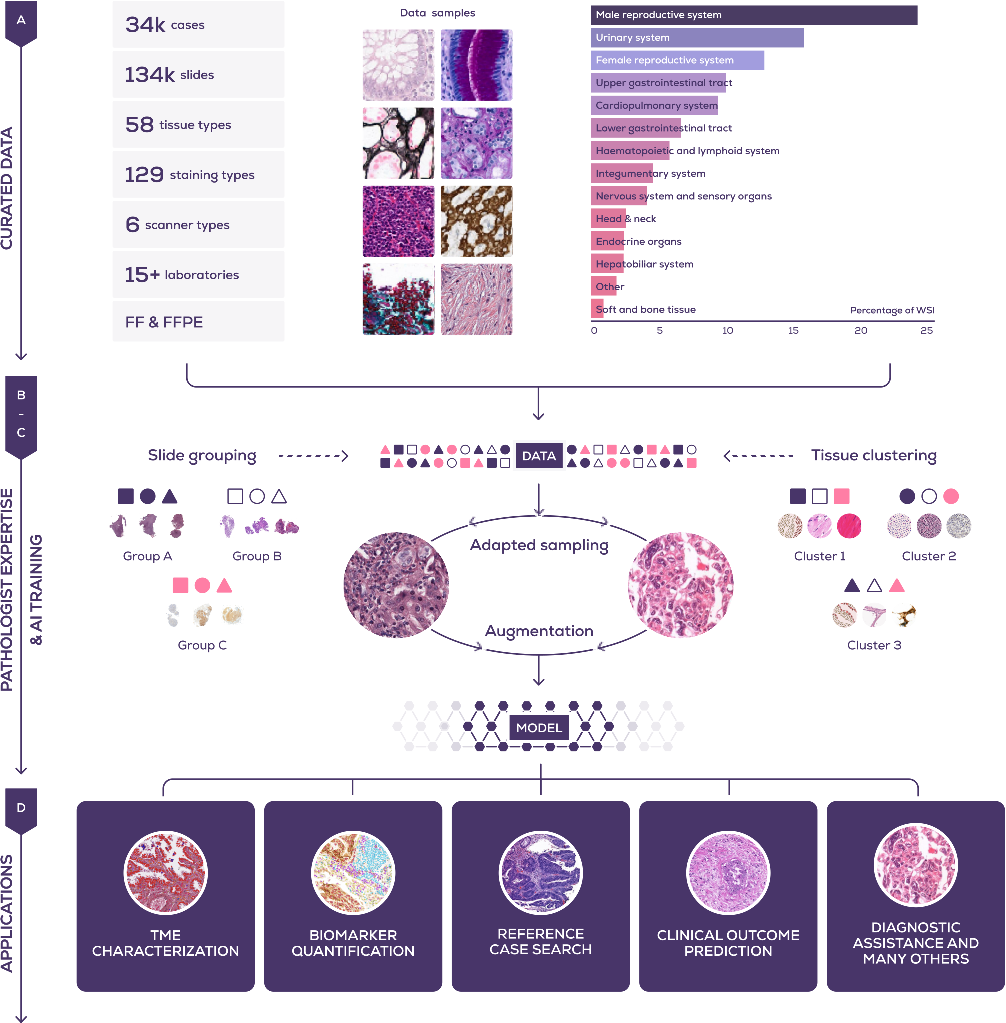}
    \caption{\textbf{Overview of the approach.} \textbf{(A) Curated data:} A dataset of 134k slides comprising 34k cases was assembled with the aim to maximize diversity while keeping size tractable. \textbf{(B) Combining computational and pathologist expertise:} Pathologists and computational scientists collaborated to group similar slides and cluster morphologically similar tissue in order to guide the data balancing in step (C).
    Based on a sample's lab of origin, tissue type, diseases, and staining modality, all slides were assigned to one of 31 groups following the principle of maximizing homogeneity within groups and heterogeneity across groups and 
    9 distinct, human-interpretable tissue clusters were formed by aggregating 100 precomputed image clusters.
    For group and cluster details see also Figure \ref{fig:data-sampling}. \textbf{(C) AI training:} Our foundation model RudolfV was trained by adapting the DINOv2 framework to sample training data from a specific distribution derived from slide groups and tissue clusters in order to balance frequent and infrequent diseases and biologies. Additionally, augmentations were extended with stain variations. \textbf{(D) Applications:} The resulting foundation model can be used for various applications in digital pathology.}
    \label{fig:overview}
\end{figure*}

% 0. Pathology is important and histopathology is key in computational pathology.
Pathology plays a central role in clinical medicine for tissue-based diagnostics and in biomedical research as a basis for understanding mechanisms of disease.
While molecular and omics-based data complement histological assessments, the microscopic evaluation of morphological changes has persisted as a key component of pathology.
Consequently, most computational pathology work focuses on whole slide image analysis.

% 1. But there are challenges with generalization, application variety, and robustness
%     Because we cannot gather enough training data due to variability and infrequent targets.
Despite artificial intelligence (AI) having led to promising proof-of-concepts and applications (e.g., \cite{histo-review, diao2021human, raciti2023clinical, binder2021morphological, keyl2022patient}),
generalization, application variety, and robustness remain challenging and still prevent the broad translation of AI applications into clinical routine diagnostics. Moreover, limited numbers of cases available for training pose a particular challenge throughout medicine, because the distribution of disease frequencies is highly skewed with few frequent, but many infrequent or even rare diseases.
Therefore, generating sufficient labeled data to cover the full spectrum of diseases including biological and technical variations regarding histomorphology, tissue processing, staining, and scanner type is prohibitively expensive.

% 2. Self supervised pretraining can help; showed promising results in other domains.
Thus, while learning from labeled data will remain the standard for most AI use cases,
novel strategies are required to address the challenges outlined above for broader clinical application. Here pretraining neural networks on (large amounts of) unlabeled data  \cite{bommasani2021opportunities} may contribute to a viable solution. The resulting so-called ``foundation models'' can subsequently be adapted to specific diagnostic tasks by training them with potentially limited, labeled data.
Training typically relies on self-supervised learning (SSL, e.g., \cite{simclr-1,dinov2,mae,krishnan2022self}) and has shown breakthrough results in other domains when employed with an abundance of unlabeled data coupled with very large models (e.g., \cite{radford2019language,brown2020language,gpt4,tu2023towards,dinov2}).

% 3. Current state in pathology FM and how differentiate:
% Pathologist domain knowledge, data curation, new stainings, are better on benchmarks.
Previous studies have started to develop foundation models for digital pathology (e.g., \cite{ciga2022self,hipt, ctranspath, owkin, uni, virchow, prism, gigapath}).
These primarily focused on the most commonly used Hematoxylin and Eosin (H\&E) staining and data from single institutions, performed no or only simple data curation and often lack deep integration of domain knowledge, a critical aspect for efficient model development.
In this work, we present an approach that uses multi-institutional multi-stain image data and integrates pathological domain expertise into the data curation and foundation model training process.
Our study pioneers the comprehensive incorporation of pathological knowledge into foundation model development, and, for the first time, integrates immunohistochemistry (IHC) and special stainings, which are crucial for biomarker evaluation, into SSL datasets and benchmarks.
The foundation model design we propose extends beyond state-of-the-art performance across different benchmarks while exhibiting favorable robustness properties.

% 4. Insights on how we did it.
Based on the premise that optimally curated, diverse data coupled with domain knowledge improves SSL, we consequently integrated pathologist expertise in the following key aspects:
(1) Curation of a dataset of 134k slides and 34k cases covering a broad spectrum of histological samples from different fixation, staining, and scanning protocols as well as different indications and laboratories across the EU and US,
(2) grouping of semantically (i.e., histomorphologically) similar slides and tissue patches to optimize data sampling for training, and (3) application of stain-specific augmentation during training.
An overview of the approach is given in Figure \ref{fig:overview}.

% 5. How we evaluated the model and that we have explainability.
We evaluate our model on a variety of common and novel applications such as H\&E-based tumor microenvironment (TME) characterization, immunohistochemistry biomarker evaluation, and reference case search including rare oncological and non-neoplastic diseases. Our proposed approach achieved the best performance on 10 out of 12 benchmarks and 28 out of 31 datasets compared to state-of-the-art pathology foundation models, while learning meaningful representations and exhibiting favorable robustness properties.

We named our model RudolfV in honor of Rudolf Virchow, who was not only the pioneer of modern pathology \cite{virchow1863cellular, reese1998fundamentals, schultz2008rudolf, goschler2021rudolf}, but also the founder of the institute from which our labs and significant histological data for this study originate from.

\section{Results}

\subsection{Pathologist-guided and diversity-focused foundation model design}

\begin{figure*}[h!]
  \centering
    \includegraphics[width=\textwidth]{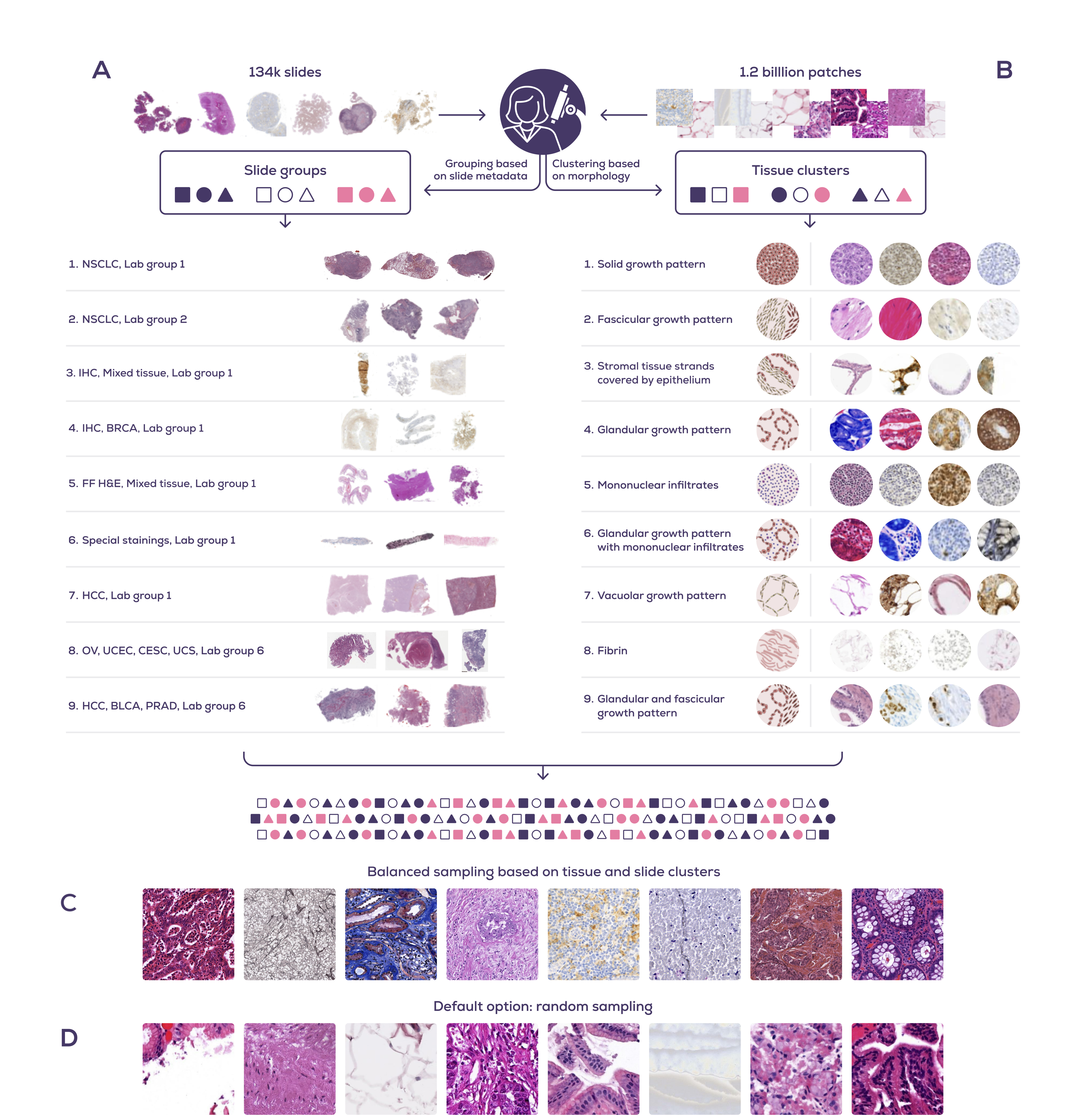}
    \caption{\textbf{Pathologist-guided and diversity-focused curation of slide groups and tissue clusters:} Pathologists and computational scientists collaborated to group slides and tissue patches. \textbf{(A)} Slides were grouped based on similarity of tissue and disease type, laboratory, and staining modality following the principle of maximizing homogeneity within groups and heterogeneity across groups. We show 9 out of the total 31 groups as examples. The full list is given in the supplement. \textbf{(B)} 1.2 billion image patches were extracted from 134k slides.
    The patches were clustered into 100 clusters and subsequently merged by pathologists into 9 morphological meaningful clusters. The first image column shows a schematic view of the morphology and other columns example morphologies.
    \textbf{(C)} The slide groups and tissue clusters were used to balance the data sampling process during training. \textbf{(D)} Example for random data sampling without balancing for comparison.}
    \label{fig:data-sampling}
\end{figure*}

In addition to the emphasis on diverse multimodal imaging data, another key aspect of the design of our foundation model is the systematic inclusion of pathological domain knowledge at crucial stages of development. While a number of self-supervised foundation modelling approaches were transferred to the domain of pathology (e.g., \cite{ciga2022self, ctranspath, owkin, uni, virchow}), so far none of these works have explicitly incorporated pathologist knowledge into dataset curation and model training, which we hypothesize to be a critical component for maximizing data informativeness and foundation model performance. Consequently, we present a novel approach that leverages domain knowledge and semi-automated data curation to generate a rich and diverse dataset that results in better foundation model learning with improved downstream task capabilities.

(1) For training RudolfV we curated a dataset of 134k slides from 34k cases. The slides were chosen by pathologists from a large archive with the objective to maximize diversity across parameters such as tissue and disease type, laboratory, staining modality, and scanner device, while keeping the dataset size tractable.  The dataset featured 133,998 slides, comprising 34,103 cases from over 15 different laboratories across the EU and US and included 58 tissue types, oncology and non-oncology diseases, 129 different histochemical and immunohistochemical staining modalities, 6 different slide scanner types, and formalin-fixed-paraffin-embedded (FFPE) and fresh-frozen (FF) samples. An overview is given in Figure \ref{fig:overview}A.

(2) After dataset curation, it is critical to present this data in a meaningful and balanced way to the model during training such that the extracted information is maximized.
This is particularly important in the medical domain, including histopathology, where the distribution of diseases and their properties are commonly heavy-tailed, i. e. there are few relatively frequent (common) diseases and a long, heterogeneous tail of less frequent diseases, both of which are relevant for comprehensive pathological assessment. Models trained only for the common diseases where enough training data is available may overlook or incorrectly classify less frequent pathologies. 
Similarly, this data imbalance is also present within a slide on a tissue level, which often contain potentially redundant properties such as large regions of normal tissue and only small areas with (rare) morphological alterations.
To this end, we propose an approach for stratifying the data on both levels through pathologist-informed slide-level grouping and patch-level clustering, which subsequently enables data sampling in a morphologically meaningful and balanced way during training. 
31 slide groups were selected by pathologists based based on slide metadata such as laboratory, tissue type, diseases, and staining modality following the principle of maximizing homogeneity within groups and heterogeneity across groups.
9 morphologically relevant tissue clusters were formed by pathologists by merging 100 precomputed tissue patch clusters.
Resulting slide-level groups and tissue patch clusters are shown in Figure \ref{fig:data-sampling}.

(3) The variety of a given dataset can be further improved by augmenting the image patches during model training. While image augmentations are standard practice in computer vision, they can be adapted in pathology via staining transformations. For training our model, for each of the 134k slides staining statistics were computed and at every training step each of the image patches was randomly augmented to resemble the staining of a different, randomly chosen other slide before presenting it to the model.

Using the above approach for data curation and model training, we found that our resulting model outperformed a model trained on a similar amount of slides (UNI \cite{uni}, 100,426 slides) on 10 out of 12 benchmarks and 27 out of 31 datasets. In addition, RudolfV outperformed a model trained on an order of magnitude more slides and twice the model size (Virchow \cite{virchow}, 1.5 million slides and 208,815 cases, 632M vs 304M parameters) on 2 out of 3 benchmarks. UNI and Virchow are both current state-of-the-art and use the same DINOv2 self-supervised learning framework \cite{dinov2} as RudolfV. The efficiency gains we achieved highlight the benefit of our pathologist-informed approach. As an additional reference, we also included Phikon \cite{owkin} in our evaluation which was trained with the DINOv2 predecessor framework iBOT \cite{zhou2021ibot} on a magnitude fewer slides (6,093 slides).

\subsection{Pan-indication tumor microenvironment characterization}

\begin{figure*}[h!]
    \centering
    \includegraphics[width=.89\textwidth]{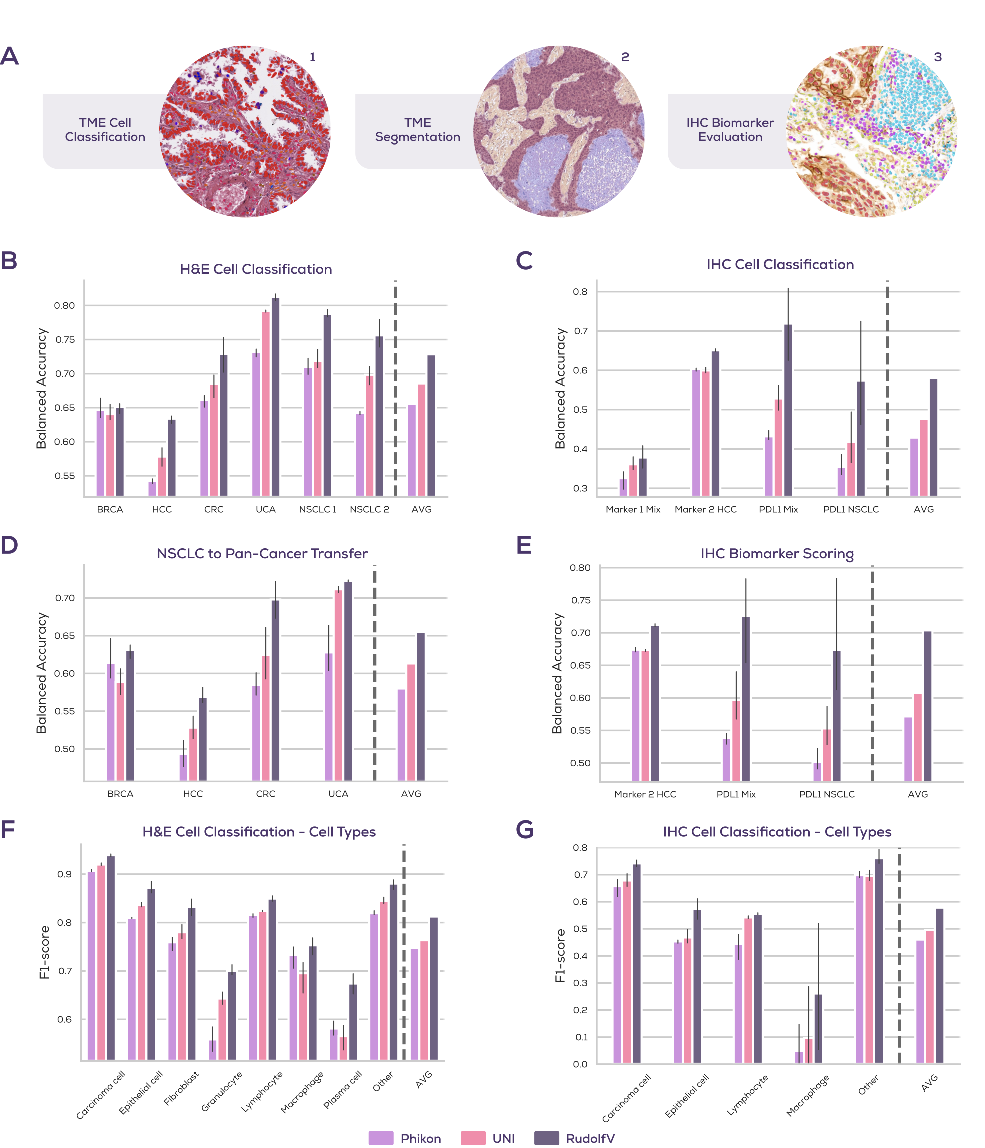}
    \caption{\textbf{TME characterization and immunohistochemistry biomarker scoring.} \textbf{(A)} Model prediction examples of H\&E TME cell classification and tissue segmentation as well as IHC biomarker evaluation. Our proposed model performed best on all benchmarks and all datasets. \textbf{(B, D, F)} Results for pan-indication H\&E TME applications. \textbf{(B)} Cell classification with 8 cell types on 5 indications. \textbf{(D)} Cell classification with 8 cell types; trained on NSCLC only and evaluated on 4 different indications. \textbf{(F)} Same as (B), but results aggregated per cell type. \textbf{(C, E, G)} Results for IHC biomarker scoring.  Our proposed approach yielded best results on all benchmarks and all datasets. \textbf{(C)} Cell classification with 5 cell types on 5 indications and 3 markers. \textbf{(E)} End-to-end membrane marker scoring for carcinoma cell, immune cells, and other cells on 4 indications and 2 markers. \textbf{(G)} Same as (C), but results aggregated per cell type.}
    \label{fig:tme}
\end{figure*}

\begin{figure*}[h!]
    \centering
    \includegraphics[width=.95\textwidth]{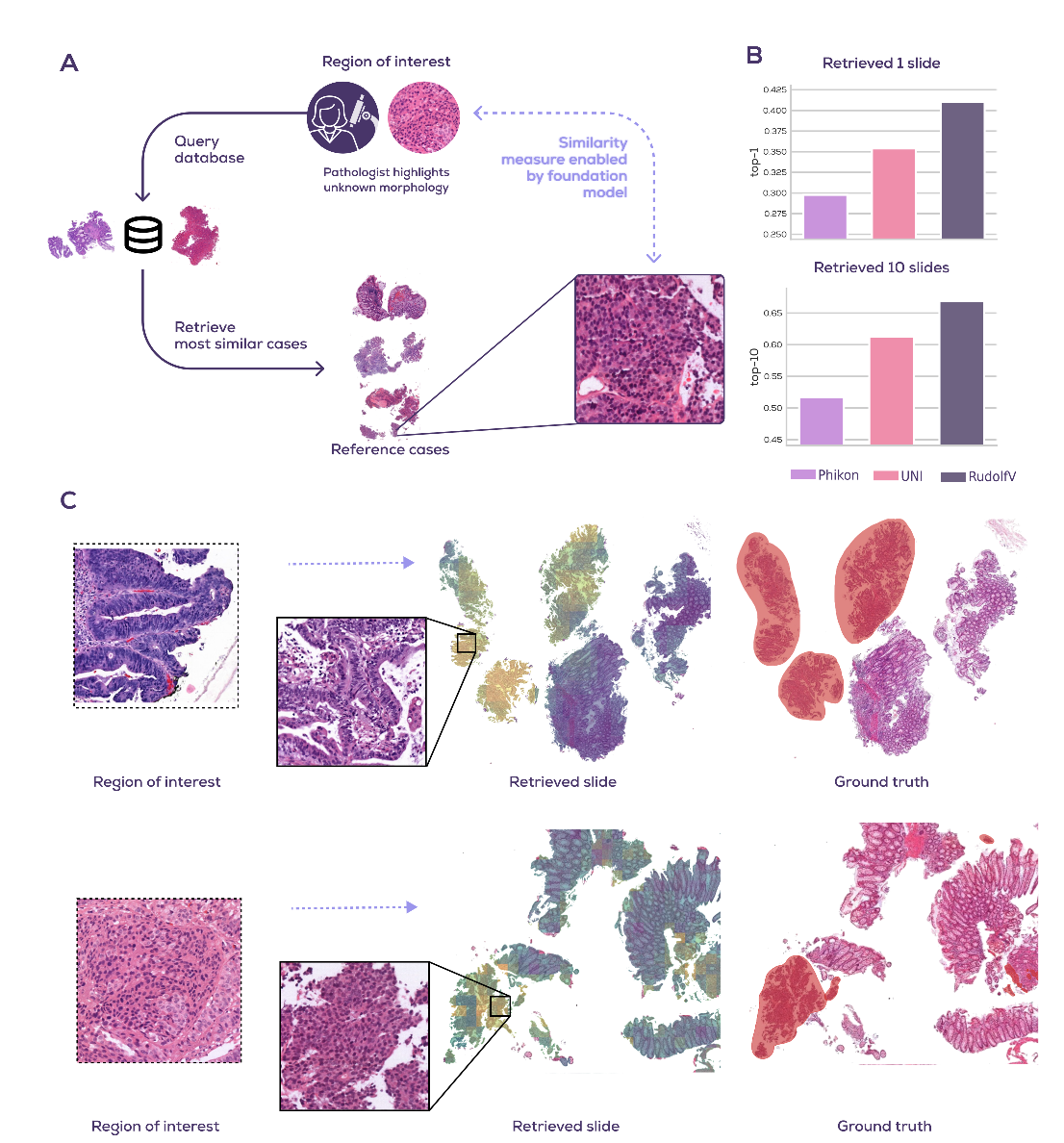}
    \caption{\textbf{Reference case search.} \textbf{(A) Workflow:} The pathologist annotates a region of interest (ROI) that is queried against a database of slides. The most similar slides are returned and shown to the pathologist allowing to consult their diagnoses. \textbf{(B) Evaluation:} Results on a benchmark with 178 rare disease slides measuring if the retrieved results contain a slide with the same diagnosis as the query slide from a database. The database contains over 6,400 slides and the rare diseases have a median occurrence of 3 or or 0.04\% in the database. The results show that in 41\% and 67\% of the queries a slide with the same diagnosis was returned when retrieving a single or 10 most similar slides respectively. For reference, not using a foundation model (ResNet-50 trained on ImageNet) yields respectively 0\% and 1.7\% correctly retrieved diagnoses. \textbf{(C) Visual aid:} The visualization of the regions with the highest similarity to the region of interest can aid pathologists and shows that the foundation model highlighted relevant morphologies. The examples are colon adenocarcinoma at the top and neuroendocrine stomach tumor at the bottom.}
    \label{fig:ref-diagnostics}
\end{figure*}

% (1) Intro to TME analysis
% (1) Importance of cell based predictions, references from Heraklion paper
% (1.a) Cell composition and locations prognostic marker
It is well known that the tumor microenvironment (TME) plays an important role in cancer progression and treatment response. Recent studies \cite{Cords2024-jf, Keren2018-ii} also show that the local spatial organization of the TME is associated with disease progression. However, the precise underlying mechanisms are only partially understood, because until recently, evaluations of large clinical cohorts were limited by semi-quantitative human assessment of the TME not covering the complete scale and spatial complexity of histological specimens.
While computational pathology has brought some remedy, only generalizing pan-tissue and pan-cancer models will allow for large-scale quantitative evaluations across primary and metastatic tumors relevant to fully assess the significance and impact of the TME on disease progression.

% (1.b) Results show that difficult task allowing for differentiation of FM results
The benchmarks used for this assessment require adapting a foundation model to characterize the TME in H\&E on a cell and tissue region level, as shown in Figure \ref{fig:tme}A1 and \ref{fig:tme}A2. Each cell should be classified into one of eight classes (carcinoma cell, epithelial cell, lymphocyte, plasma cell, macrophage, granulocyte, fibroblast, other cell) and each tissue location should be attributed to a TME area (tumor, stroma, necrosis, other). Such TME characterization goes beyond the sole detection of cancerous tissue \cite{virchow} and requires robust prediction of immune cell types and tissue regions across cancer and tissue types.
Additionally, pathology cases frequently include tissue with the same disease but taken from various origins. This was modeled in the benchmarks by including data from 5 indications and 8 organs.

% (2) H&E cell classification and segmentation
% (2.b) [Result] H&E pan-indication cell classification linear probing
Figure \ref{fig:tme}B shows results for H\&E cell classification. RudolfV performed best on all indications with an average improvement of 6.3\% balanced accuracy.
When considering the performance per cell type (Figure \ref{fig:tme}F), it can be observed that our proposed model also shows best performance for all cell types. Especially for immune and stromal cells such as fibroblasts, granulocytes, macrophages, and plasma cells the model improved by 10.8\% on average over the closest contender.
% (2.c) [Result] H&E pan-indication segmentation frozen encoder
For TME area segmentation the results given in Figure \ref{fig:learning}C show that our approach worked favorably with an average improvement of 0.45\% F1-score.

% (3) Pan-indication transfer capabilities
% (3.a) [Result] H&E NSCLC trained and evaluated on all linear probing
Foundation models in pathology are trained on a wide range of tissue types and diseases and it is assumed that they learn a cross-tissue disease understanding. We examined this in an experiment where a TME cell classification model was only trained on non-small cell lung cancer (NSCLC) data and evaluated on other indications. The results given in Figure \ref{fig:tme}D and in the supplement indeed show that foundation models were able to generalize to new indications and that not using a pathology foundation model lead to significantly worse performance (36.2\% without foundation model vs.\ 65.5\% balanced accuracy with RudolfV) underlining the utility of domain-specific foundation models and that such models indeed learn cross-tissue and -indication patterns.

\subsection{Pan-indication immunohistochemistry biomarker scoring}
% (1) IHC cell classification and scoring
Histomorphological assessment of tissue specimens stained with H\&E is the basis of pathological diagnostics. At the same time, almost all cancer cases require a growing number of additional immunohistochemical stainings for tumor subtyping and precision biomarker assessment. The scoring of a biomarker typically entails two steps: first to differentiate the cell types for which the marker should be evaluated and second to quantitatively assess its expression. Our benchmarks are split accordingly. The first requires learning to differentiate key cell types (carcinoma cell, epithelial cell, lymphocyte, macrophage, other cell) in different indications and for different biomarkers. The second requires learning to score the marker positivity in the membrane compartment. An example model output is given in Figure \ref{fig:tme}A3.

% (2) [Result] IHC pan-indication cell classification linear probing
Figure \ref{fig:tme}C shows results for cell type characterization across PD-L1 and two proprietary markers in 5 indications. Our model performed best on all evaluation sets and with an average improvement of 21.6\% balanced accuracy. When considering the performance per cell type in Figure \ref{fig:tme}G the model performed particularly well on the identification of epithelial cells and macrophages.
% (3) [Result] Membrane positive PD-L1+marker 2 out of sample end-to-end scoring
When taking membrane marker evaluation for the cell types of interest (carcinoma cells, immune and other cells) across PD-L1 and one other membrane marker in 5 indications into account our proposed approach showed an average improvement of 15.8\% compared to the next best foundation model and showed best results for each evaluation set.

\subsection{Reference case search}
% (0) Defend reference diagnostic task
It is common clinical practice to consult disease experts for particularly hard or rare cases. In pathology a second opinion is typically obtained by sending such cases to a reference pathologist. Foundation model-powered applications may reduce the need to consult expert pathologists by foundation model-powered database searches that retrieve histologically similar cases \cite{yottixel, sish, retccl, hshr}. The workflow is depicted in Figure \ref{fig:ref-diagnostics}A.

Image-based search in a histological database requires a semantically meaningful representation of tissue. So far, it has been challenging to automatically generate sufficiently meaningful tissue descriptions. Pathology foundation models alleviate this shortcoming as they have seen huge corpora of cases and thereby developed a strong ``understanding'' of histology. To match infrequent diagnoses correctly, it is particularly important that the foundation model has seen a great variety of data, including rare diseases.

% (1) Reference diagnostic of rare cancer types on a dataset of ~5k cases
The following benchmark provides evidence that foundation models will be able to support pathologists in the described scenario. Specifically, we have compiled a database of over 6,400 cases from more than 8 tissue types and from over 44 different cancers and non-oncological diseases including 178 cases from 34 rare diseases of colon or stomach. The rare cases have a median occurrence of 3 or 0.04\% in the database and examples are non-invasive neoplasms, various malignant solid or haematological neoplasms, including metastatic melanoma, neuroendocrine tumors and marginal zone lymphoma, and infectious diseases such as helminthiasis. For the 178 rare cases a board-certified pathologist highlighted a region of interest indicating an ``unknown'' disease.
The image representation of the foundation model was then used to search histological slides with similar patterns in the database.

% (1a) [Result] Colon and stomach reference diagnosis results
% (2) Stress that very hard task allowing for more granular FM comparison
While, as expected, the task remains difficult, the results in Figure \ref{fig:ref-diagnostics}B show that foundation models clearly enable such applications by retrieving cases with the same rare diagnosis for 41\% of the queries when returning a single case and 67\% of the queries when returning 10 cases. For reference, not using a foundation model yielded 0\% and 1.7\% correct diagnosis when returning a single or 10 cases, respectively. We can also observe that RudolfV exhibited the best performance with an average improvement of 16\% and 9\% for single and 10 case retrieval, respectively, indicating that the representations learned contain information about (rare) cancers. 

Figure \ref{fig:ref-diagnostics}C shows the visual aid presented to the querying pathologist. The visualization guides the pathologist to the areas most similar to the region of interest used in the query. Such visualizations facilitate the interpretability and provide reassurance that the foundation model indeed recovers the relevant slide regions.

\begin{figure*}[h!]
    \centering
    \includegraphics[width=.95\textwidth]{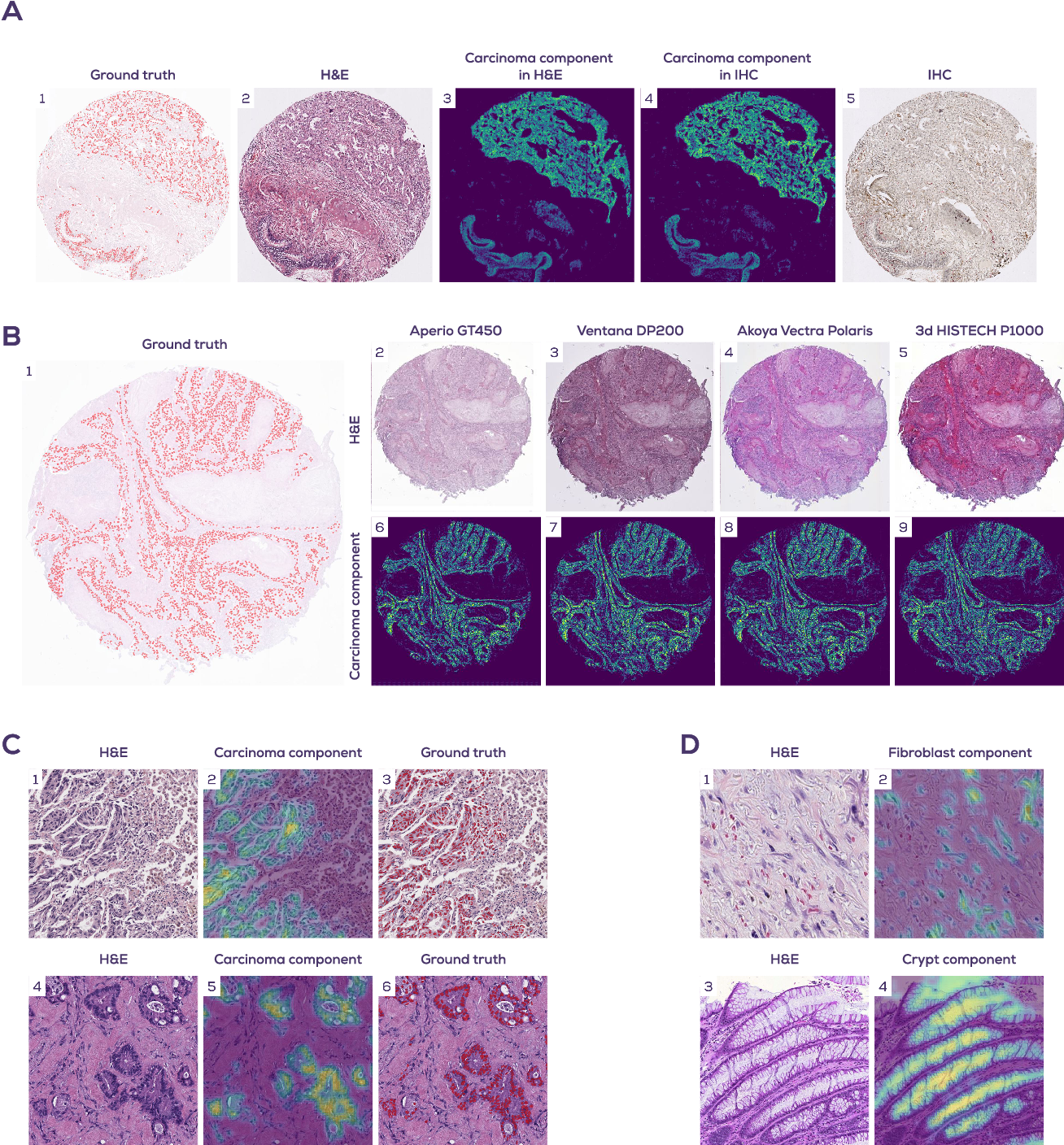}
    \caption{\textbf{Foundation model characteristics and robustness properties.} Foundation models learn concepts from data without human supervision. Learned concepts can be examined via principal components~\cite{PCAShlens2009,PCAhotelling1933analysis,dinov2,virchow} of the embedding space, which we qualitatively analyzed for commonalities and robustness in different settings. \textbf{(A) Pan-Staining:} The \textit{same} tissue was stained in HE (A2) and IHC (A5). The principal component visualization (A3 and A4) highlights the carcinoma component, which is consistent across stains and approximately overlaps with the ground truth output (A1) of a supervised carcinoma detection model. \textbf{(B) Pan-Scanner:} The same tissue was scanned by 4 different scanners (B2-B5). Despite very different visual appearance of the scans, the carcinoma component (B6-9) is consistent across scanners and approximately overlaps with the ground truth (B1), indicating a high degree of scanner invariance of the learned representation. \textbf{(C) Detailed view:} A detailed view of the carcinoma component, showing that most carcinoma cells are covered by the (self-supervised) learned representation. \textbf{(D) Additional components:} Additional components such as fibroblasts or crypts are identified by the foundation model.}
    \label{fig:xai}
\end{figure*}

\begin{figure*}[h!]
    \centering
    \includegraphics[width=\textwidth]{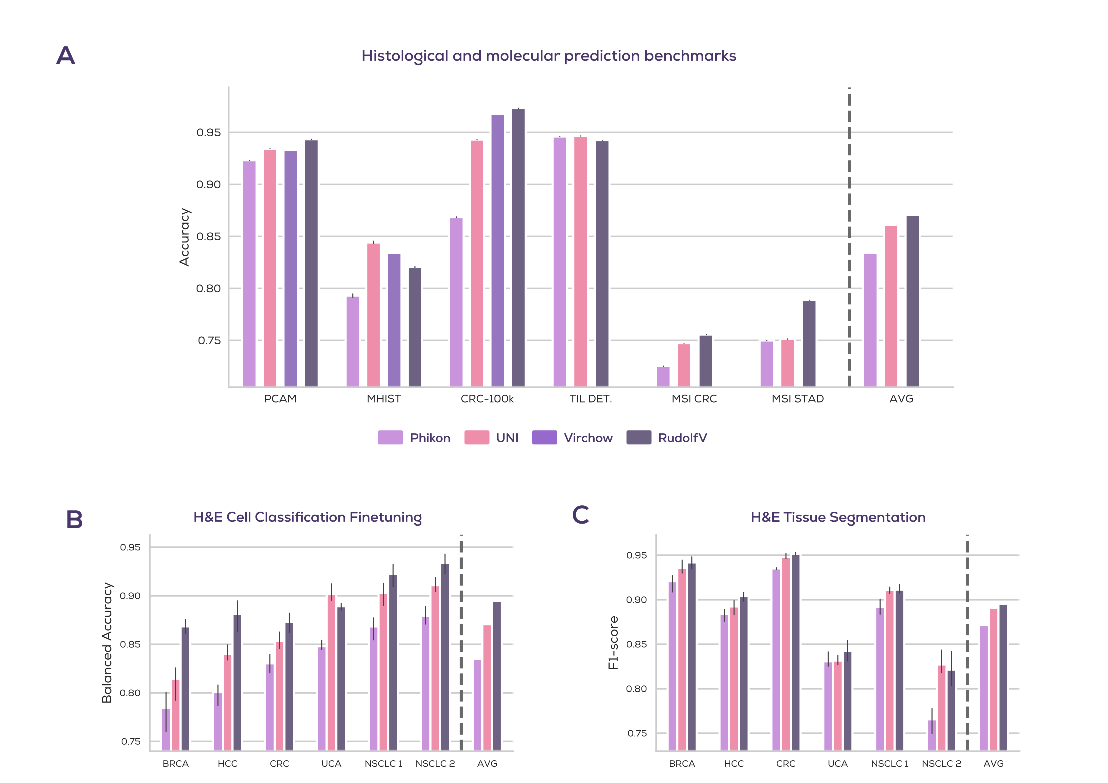}
    \caption{\textbf{Histological and molecular prediction benchmarks.} 
    \textbf{(A) Benchmark results:} RudolfV outperformed other foundation models on 4 out of 6 tasks as well as on average. Results of Virchow \cite{virchow} were only available for the first 3 benchmarks. 
    \textbf{(B) Finetuning results:} Cell classification results per dataset when also optimizing the foundation model encoder next to the linear classification head, in constrast to Figure \ref{fig:tme}B, where only the linear classification head was optimized. RudolfV yielded best results compared to competing foundation models on all but one dataset and on average.
    \textbf{(C) H\&E TME segmentation results:} H\&E pan-indication tissue segmentation results with 4 tissue classes and on 5 indications. RudolfV performed favorably on average and on all but one dataset.
    }
    \label{fig:learning}
\end{figure*}

\subsection{Foundation model characteristics and robustness}
The previous sections indicate that our proposed foundation model learned meaningful histological representations. In this section, we explore which concepts are learned by RudolfV as well as qualitatively assess the model's robustness to batch effects and therefore its capability to generalize across variations encountered across hospitals and diseases in routine diagnostics.

The data representation produced by a foundation model should be semantically meaningful, i.e.\ in our case represent morphological concepts. We can explore learned concepts by examining the most important data dimensions produced by the foundation model and, as in representation learning linear disentanglement of concepts is typically desired \cite{bengio2013}, examine the principal components~\cite{PCAShlens2009,PCAhotelling1933analysis,dinov2,virchow} in foundation model embedding space. 
In Figure \ref{fig:xai}A1 and A2 carcinoma cells and the according H\&E image are shown and in A3 the corresponding carcinoma principal component produced by the foundation model is visualized. One can observe that the carcinoma component approximately segments the tumor area.

RudolfV was trained with data from a variety of staining types and it is hypothesized that cross-staining concepts are learned by the model. We investigated this by considering a second scan of the {\it same} tissue after re-staining it with an MPO + CD68 IHC staining (Figure \ref{fig:xai}A5). From Figure \ref{fig:xai}A3-4 one can observe that the model approximately identified the same cancer regions in both stainings, indicating that the model has learned a morphological representation that tends to be invariant towards staining variations.

A related hypothesis is that foundation models are robust to batch effects, such as lab and scanner variations. We examined this by observing the concepts learned for the same tissue scanned with four different scanner types. Figure \ref{fig:xai}B shows that the representation of our model are approximately the same across scanners, indicating robustness to scanner variations.

Figure \ref{fig:xai}C-D gives a closer look on the granularity and different scales of the learned representation. The results shows that tumor regions (Figure \ref{fig:xai}C1) as well as fibroblasts (Figure \ref{fig:xai}D) and crypts can be identified on granular level by our foundation model.

It is remarkable that the identification of these different histological patterns happens in an unsupervised manner and thus without any specific pathologist input apart from the general guidance by pathologists at the initial stage of the foundation model design.

\subsection{Histological and molecular prediction benchmarks}
% (1) Tile-based benchmarks
% [Result] External benchmarks
Figure \ref{fig:learning}A shows results on additional benchmarks that include both histological classification and molecular prediction tasks. These analyses show that our proposed model was on average the best performing model. For the first three benchmarks, Virchow \cite{virchow}, a model trained on an order of magnitude more data, was a close contender highlighting again the efficiency of our pathologist-curated approach.

% (3) FM-initialization matters for finetuning as well
% [Result] H&E pan-indication cell classification linear probing
Many benchmarks assume that the learned embeddings of a foundation model cannot be modified, focusing on the evaluation of the learned representation. However, in practice one can further finetune the parameters of the foundation model to improve on the task at hand and it is hypothesized that initializing a downstream model with pretrained model weights is beneficial for finetuning, as the foundation model incorporates prior knowledge about a given domain. Figure \ref{fig:learning}B shows that the choice of foundation model used for initialization strongly matters to the task at hand. Our proposed model achieved the best performance on all tasks and showed an average relative improvement over not adapting the foundation model of 22.9\%. Thus, the intrinsic parameter structure of a model heavily influences subsequent learning.

\section{Discussion}
% (1) Summary, justification, comparison. Mention orthogonal to other approaches and can be combined.
In this work we presented the foundation model RudolfV, whose design incorporates medical expert knowledge into data curation and model training with an emphasis on data diversity. This approach allows for efficiency gains in foundation model training and performance beyond current state-of-the-art~\cite{uni, virchow, gigapath, ctranspath, prism, owkin}. The study evaluates a wide range of established benchmarks and novel applications, showing that RudolfV performed best in 10 out of 12 benchmarks and 28 out of 31 datasets among state-of-the-art pathology foundation models.
Our results suggest that leveraging pathologist domain knowledge and data diversity can have a similar effect as an order of magnitude more data available for training and we believe that foundation models built according to the design principles proposed here will facilitate the inclusion of the clinically highly relevant broad spectrum of less frequent and rare diseases encountered in real-world deployments. To this end, training on very large and highly diverse as well as curated and balanced datasets built with expertise knowledge will likely be necessary.

% (2) Discussion of results and what went well. Precision downstream applications not covered so far. Non-H&E results. Learning based results.
The curated development of RudolfV features a dataset of significant diversity, including data from over 15 laboratories, 58 tissue types, 129 different stainings, and 6 scanner types.  Preprocessing required the extraction of 1.2 billion image patches and the close collaboration between computational scientists and pathologists in order to develop a slide grouping and tissue clustering strategy that optimized foundation model training. Recent work in computer vision (e.g., \cite{dinov2, vo2024automatic}) has shown similar benefits of using domain knowledge for curating diverse datasets for SSL pretraining on natural images. To evaluate our approach we integrated public benchmarks and developed novel applications focusing on tissue-centric tasks such as H\&E TME characterization, IHC biomarker scoring, and reference case search. This is the first work to develop and compare pathology foundation models on IHC benchmarks and to incorporate a reference case search task including oncology and non-oncology data as well as rare diseases. On both benchmarks RudolfV performed favorably indicating the benefit of multi-stain data as well as the data curation approach.
Furthermore, our qualitative analysis indicates robustness of the proposed model RudolfV across different stainings and scanner types. For supposedly easier tasks such as tissue segmentation, we observed a performance plateau among foundation models.
We further showed that the choice of foundation model also significantly influenced the performance on finetuning tasks where foundation model parameters are adapted to a downstream task.

% (3) Limitations: of data, of pretraining method, not slide level and not text, not multi-scale (but evaluated on several scales).
Despite the strong performance of our foundation model, our study has some limitations. In spite of its vast diversity, our training dataset and benchmarks lack cytopathology and hematopathology cases. Future research should explore how the relationship of data diversity and scale evolves with more extensive datasets and larger model sizes. Due to its proven performance and wide adoption as well as to enable comparability to other work, this study uses the DINOv2 framework \cite{dinov2}. In recent computer vision and computational pathology work \cite{i-jepa, pluto} novel pretraining methods are explored and might have a substantial impact on the performance and data sensitivity of foundation models. A similar reasoning applies to the choice of the backbone network architecture (ViT \cite{vit}) and the use of multi-scale data during pretraining where novel methods \cite{vmamba, medmamba} and studies \cite{pluto}, respectively, have been introduced concurrently to this work. Integration with slide-based foundation model concepts \cite{hipt,longViT,prism} as well as extensions to multiple modalities, e.g.\ text and vision \cite{histogpt, prism} or OMICS and vision, are promising avenues for future investigation.

\renewcommand*{\bibfont}{\normalfont\small}
\printbibliography

\subsection*{Acknowledgements}

We would like to thank the teams at Aignostics for supporting this work in many ways. Without their ground work this work would not have been possible. Special thanks goes to Melina Kaplan and Jesús González for their support with the graphics and Todd Dembo for his editing support.

The results shown here are in whole or part based upon data generated by the TCGA Research Network: https://www.cancer.gov/tcga.

This work was in part supported by the German Ministry for Education and Research (BMBF) under Grants 01IS14013A-E, 01GQ1115, 01GQ0850, 01IS18025A, 031L0207D, and 01IS18037A. K.R.M. was partly supported by the Institute of Information \& Communications Technology Planning \& Evaluation (IITP) grants funded by the Korea government (MSIT) (No. 2019-0-00079, Artificial Intelligence Graduate School Program, Korea University and No. 2022-0-00984, Development of Artificial Intelligence Technology for Personalized Plug-and-Play Explanation and Verification of Explanation).

Correspondence to Maximilian Alber (max@aignostics.com), Frederick Klauschen (frederick.Klauschen@med.uni-muenchen.de), and Klaus-Robert Müller (klaus-robert.mueller@tu-berlin.de).

\section{Methods}

\subsection{Pathologist-guided and diversity-focused foundation model design}

\paragraph{Data curation}
We curated a dataset focusing on data diversity both with respect to medical and technical aspects. The data covers a broad range of disease entities and was generated with different fixation, staining, scanning protocols, and in different labs across the EU and US.
The dataset details are as follows:

\begin{itemize}
    \item \textbf{Cases and slides:} The dataset consists of 133,998 slides comprising 34,103 cases. 108,433 (81\%) slides have been chosen from our archive sourced from different labs and 26,565 (19\%) slides are from The Cancer Genome Atlas (TCGA).
    \item \textbf{Tissue location:} The tissue originates from 14 organ systems (Figure \ref{fig:overview}) comprising 58 different tissue types (examples for the granularity are lung, heart, adrenal gland, pancreas, oral cavity, bone marrow).
    \item \textbf{Labs:} The slides were created in over $15$ different labs in the EU and US.
    \item \textbf{Staining:}  Our dataset contains 3 broad staining categories: H\&E (68\%), IHC (15\%), and other (17\%). Subdividing each staining category further results in 129 unique staining types. Examples are H\&E, ER, PR, PD-L1, HER-2, Giemsa, PAS, Gomori, and others.
    \item \textbf{Scanning:} 6 different scanner types were used to scan the slides, including popular scanners such as Roche Ventana DP600, Leica Aperio GT 450, and 3DHISTECH PANNORAMIC 1000. Scanning magnifications were typically 20x, 40x, and 80x.
    \item \textbf{Preparation:} The dataset contains both FFPE and FF tissue samples. 
\end{itemize}  

From each slide we extract patches within given tissue boundaries derived from an in-house model. The patch size is $256\times256$ pixels at $0.5$ mpp. This results in $1.25$ billion image patches in total for further processing and training.

\paragraph{Data sampling}
Histology slides typically come with metadata available for data curation. 
Based on a sample's lab of origin, tissue type, diseases, and staining, all slides were assigned by pathologists and computational scientists to one of 31 groups following the principle of maximizing homogeneity within groups and heterogeneity across groups. Example groups are visualized in Figure \ref{fig:data-sampling} and all groups are described in the supplement.
Weights were assigned to the individual groups, which were then used to sample patches during training. Groups with non-H\&E stainings were upsampled to increase overall heterogeneity.

In contrast, individual tissue patches typically have no patch-specific metadata available.
For optimizing patch sampling during training, we therefore created clusters based on computer vision image features and pathologist expertise.
First, we extractd 36 standard computer vision image features for all $1.25$ billion image patches. These include image statistics such as mean and standard deviation for individual channels in common color modes such as RGB, LAB, HSV, and---tailored towards digital pathology---HED \cite{ruifrok2001quantification}. In contrast to \cite{dinov2}, we did not use deep network features to reduce the computational burden. Next, we further reduced computational overhead by always sub-sampling 500 patches for each slide and perform k-means clustering ($k=100$) on the respective image features. We propagated the labels of the clustering via the k-nearest-neighbor algorithm to assign a cluster label to each of the $1.25$ billion patches. Finally, board-certified pathologists filtered and aggregated the 100 image clusters, assigning a description and a sampling probability (importance weight) to each combined cluster. This resulted in 9 distinct human-interpretable tissue clusters and $1.2$ billion image patches after filtering out irrelevant clusters. The clusters are visualized in Figure \ref{fig:data-sampling} and are described in the supplement. During training, we sampled patches according to the sampling probability of the clusters.

\paragraph{Data augmentation}
In pathology it is known that staining and scanning outputs vary between labs and even within the same lab over a given period of time. Consequently, in histopathology studies, staining and scanner information can produce spurious correlations and so-called ``Clever Hans'' effects \cite{clever-hans} when correlated with label information \cite{histo-review}. To address this shortcoming, we transferred and augmented stain and scanner color profiles between patches in addition to the standard color augmentations in the view generation process of DINOv2 \cite{dinov2}. For each view, we picked a random other patch in the batch and transferred the patch color profile to the slide color statistics of the selected patch \cite{reinhard2001color}. This discourages the model from exploiting staining and scanner color features for learning representations.

We further added 90 degree rotations as well as horizontal and vertical flipping to the augmentations in DINOv2, incorporating the prior that objects on histopathological slides have no canonical orientation. Following \cite{faryna2021tailoring, kang2023benchmarking}, we removed the solarization augmentation from the DINOv2 standard augmentations. 

\paragraph{Pretraining}
We built upon the official DINOv2 implementation \cite{dinov2} with registers \cite{vit-registers} and defined the final foundation model embeddings following \cite{virchow}.
As \cite{remedis} has shown that self-supervised histopathology models can benefit from pretraining on natural images, we initialized the student model with the distilled DINOv2 version pretrained on LVD-142M \cite{dinov2}. This has shown to improve performance in smaller experimental setups. We trained a ViT-L/14~\cite{vit} with a batch size of 960 on 16 A100-40GB GPUs with a cosine learning rate schedule (from $2 \times 10^{-4}$ to 0), 100k warm-up steps and a weight decay schedule from $0.04$ to $0.2$ for 625k iterations.

\subsection{Evaluation Setup}
We evaluated the model on a total of 12 benchmarks: H\&E cell classification, H\&E tissue segmentation, H\&E pan cancer transfer, IHC cell classification, IHC biomarker scoring, reference case search, PCAM, MHIST, CRC-100K, TIL DET, MSI CRC, and MSI STAD. In this section, we provide details for the different evaluation protocols and data. All cases and data used for training RudolfV has been separated from the evaluation data. For the histological and molecular prediction benchmarks see additional remarks in Section \ref{sec:methods:eval:additional}.

\subsubsection{Pan-indication tumor microenvironment characterization}\label{sec:methods:eval:TME}

\paragraph{H\&E cell classification} For the experiments in Figure \ref{fig:tme}B,D,F and Figure \ref{fig:learning}B we performed 8-class cell classification tasks predicting carcinoma cell, epithelial cell, plasma cell, fibroblast, lymphocyte, granulocyte, macrophage, or other cell. A single classification layer on top of the final foundation model embedding output by the frozen, pretrained image encoder were optimized using the ADAM optimizer \cite{adam}. The corresponding benchmark data is based on 123,301 annotated training and 55,706 evaluation image patches extracted with $128\times128$ pixels at 0.5 mpp resolution. Cases were separated into training and evaluation data. Annotations were gathered by board-certified pathologists on 347 slides of 8 different tissue types (bladder, breast, colon, kidney, lung, lymph node, peritoneum, soft tissue) and 5 different indications (urothelial cancer (UCa), breast cancer (BRCA), colorectal cancer (CRC), hepatocellular carcinoma (HCC), and non-small-cell lung cancer (NSCLC)). Each indication is represented in the evaluation data with a dedicated hold-out set. `NSCLC 2' indicates an additional hold-out set from a different lab group. To allow for a fair comparison among all benchmarked foundation models, we sweeped over both learning rate and weight decay in the same fixed range. We always selected the model which performed best on the validation part of the train set for evaluation. For each model the experiment was repeated 3 times with different random seeding. In Figure \ref{fig:tme}B we report balanced accuracy which accounts for class imbalances. Individual cell type metrics in Figure \ref{fig:tme}F are reported using F1 scores, combining precision and recall per class.

\paragraph{H\&E tissue segmentation} The experiment in Figure \ref{fig:learning}C is a pixel-wise 5-class semantic segmentation task predicting carcinoma, stroma, epithelial tissue, necrosis, or other tissue. To this end, we optimized a decoder head acting on the output of the frozen, pretrained foundation model encoder. The decoder head translates the foundation model output to the original input image resolution using a series of convolutional layers and upsampling operations, similar to \cite{kirillov2023segment}. The corresponding benchmark data is based on 48,924 annotated training patches and 21,416 evaluation patches extracted with $224\times224$ pixels at 0.5 mpp resolution. Cases were separated into training and evaluation data. Annotations were gathered by board-certified pathologists on 641 slides of 8 different tissue types (bladder, breast, colon, kidney, lung, lymph node, peritoneum, soft tissue) and 5 different indications (urothelial cancer (UCa), breast cancer (BRCA), colorectal cancer (CRC), hepatocellular carcinoma (HCC), and non-small-cell lung cancer (NSCLC)). Each indication is represented in the evaluation data with a dedicated hold-out set. `NSCLC 2' indicates an additional hold-out set from a different lab group. We evaluated the model which performed best on the validation set split from the training data. For each model the experiment was repeated 3 times with different random seeding.

\subsubsection{Pan-indication immunohistochemistry biomarker scoring}

\paragraph{IHC cell classification}
For the experiments in Figure \ref{fig:tme}C,G we performed a 5-class cell classification task predicting carcinoma cell, epithelial cell, lymphocyte, macrophage, or other cell. A single classification layer on top of the final foundation model embedding output by the frozen, pretrained image encoder was optimized using the ADAM optimizer~\cite{adam}. The corresponding benchmark data is based on 16,658 annotated training and 15,804 evaluation image patches extracted with $128\times128$ pixels at 0.5 mpp resolution. Cases were separated into training and evaluation data. Annotations were gathered by board-certified pathologists on 122 slides of 4 different indications (urothelial cancer (UCa), breast cancer (BRCA), hepatocellular carcinoma (HCC), and non-small-cell lung cancer (NSCLC)). The benchmark uses one nuclear and cytoplasmic marker `Marker 1', one membrane marker `Marker 2', and the marker PD-L1. Each biomarker is represented in the evaluation data with a dedicated hold-out set with `PD-L1 Mix' and `Marker 1 Mix' indicating mixed indication holdout sets. Moreover, `PD-L1 NSCLC' indicates an additional hold-out set from a different lab group. To allow for a fair comparison among all benchmarked foundation models, we sweeped over both learning rate and weight decay in the same fixed range and evaluated the models which performed best on the validation set split from the training data. The evaluation performance is reported using standard cell classification metrics. For each model the experiment was repeated 3 times with different random seeding. In Figure \ref{fig:tme}C we report balanced accuracy to account for eventual class imbalance. Individual cell type metrics in Figure \ref{fig:tme}G are reported using F1 scores, combininging precision and recall per class. 

\paragraph{IHC scoring}
For end-to-end membrane marker scoring in Figure \ref{fig:tme}E, we complemented the IHC cell classification task with a binary classification task predicting positive or negative cell membrane biomarker expression for a center cell in a given image patch. During training we optimized a single classification layer on top of the final foundation model as well as the foundation model encoder using the ADAM optimizer \cite{adam}. The membrane scoring model was trained on 8,358 annotated training image patches extracted with $128\times128$ pixels at 0.5 mpp resolution. For evaluation we used 10,126 image patches for which we collected both biomarker positivity and 5-class cell type annotations as for the IHC cell classification benchmark. Cases were separated into training and evaluation data. Annotations were gathered by board-certified pathologists on 141 slides of 4 different indications (urothelial cancer (UCa), breast cancer (BRCA), hepatocellular carcinoma (HCC), and non-small-cell lung cancer (NSCLC)). Each bimarker is represented in the evaluation data with a dedicated hold-out set with `PD-L1 Mix' indicating a mixed indication holdout set. `PD-L1 NSCLC' indicates an additional hold-out set. To allow for a fair comparison among all benchmarked foundation models, we sweeped over both learning rate and weight decay in the same fixed range and evaluated the model which performed best on the validation set split from the training data. End-to-end membrane scoring was then formulated as a combined task of predicting the cell type (using the IHC cell classfication models in Figure \ref{fig:tme}C) and the biomarker expression of each cell. For performance assessment we pooled the overall set of combined prediction classes into membrane positive carcinoma cell, membrane negative carcinoma cell, membrane positive immune cell, membrane negative immune cell, and a general other cell category to reflect clinical practice. Performance is reported as balanced accuracy over the categories. For each model the experiment was repeated 3 times with different random seeding.

\subsubsection{Reference case search}
To evaluate our reference diagnostics use case, we used 6,428 slides from more than 8 tissue origins including over 44 different cancers and non-oncological diseases. 178 slides of rare GI-biopsies (94 colon and 84 stomach) were annotated with regions of interest (ROI) that show the characteristics of the disease and have a median disease occurrence of 3 or or 0.04\% in the database. A database with representations of all patches from each slide was compiled. To query a slide, the representations for the patches highlighting the region of interest were considered. For all of the ROI patches, the cosine similarity to all patch representations of slides in the database was calculated. The average similarity of the top-k most similar patches from a slide was considered as score for a ROI patch. The mean similarity score across ROI patches was used as final similarity measure between the query slide and the candidate slide. After computing a slide similarity score for each candidate slide, we check whether a slide with the same diagnosis as the query slide is in the top-k results highlighting that a relevant case was found for the pathologist. We report top-k accuracy for $k = 1, 10$ in Figure \ref{fig:ref-diagnostics}.

% (3) Make the differences to prototype based classification clear.
It is worth noting that this task is different to few-shot learning and prototype-based applications \cite{snell2017prototypical, wang2019simpleshot, uni} in that no learning or prototype selection is involved and the representations are used as is. Furthermore, our evaluation is focused on the utility of foundation model representations for rare disease retrievals and could be combined with the methods in \cite{yottixel, sish, retccl, hshr}.

\subsubsection{Foundation model characteristics and robustness}
Following \cite{dinov2, virchow} the features learned by RudolfV are visualized using Principal Component Analysis (PCA) \cite{PCAShlens2009,PCAhotelling1933analysis}. We decomposed the foundation model embeddings of different sets of patches and visualize one of the first components selected by a pathologist. The mean vector of all embeddings at the same location was subtracted to remove the positional encoding information before performing PCA. 
To make the heatmaps clearer, we only visualize the positive component of each axis of interest.

In order to evaluate consistency of the foundation model representation across different scanners and stains, different views of the same physical tissue are used.
For the views of different scanners the tissue was re-scanned with four different scanners (Aperio GT450, Ventana DP200, Akoya Vectra Polaris, and 3d HISTECH P1000). For the views of different stainings tissue was stained with H\&E and then re-stained with IHC. In both cases the different images were co-registered pixel accurate with an in-house registration algorithm.

\subsubsection{Histological and molecular prediction benchmarks}
\label{sec:methods:eval:additional}

\paragraph{Histological and molecular prediction benchmarks} Additional experiments were performed on histopathology benchmarks with linear probing. During linear probing, we resized the images to 224 pixels and fit a linear classifier to the embeddings. We used the ADAM optimizer \cite{adam} with learning rate $5 \times 10^{-5}$, batch size 128, and cosine learning rate schedule. For each model the according experiment was repeated 5 times with different random seeding. In the following, we give details about the considered datasets.

PCAM, MHIST, CRC-100K benchmarks are an out-of-distribution validation for our model as the data was generated by separate labs. For TIL Det., MSI CRC, MSI STAD there is a potential overlap between pre-training and task data, but we assume the impact to be negligible. Please note, similarly for Phikon \cite{owkin} it is unclear if pre-training and TIL Det., MSI CRC, MSI STAD task data overlap.

\paragraph{PCAM} The PatchCamelyon dataset consists of 327,680 H\&E images ($96\times96$ pixels at 10x magnification) extracted from scans of lymph node sections of breast cancer. Each image is annotated with a binary label indicating presence of metastatic tissue. An image is labeled as metastatic tissue when one pixel was annotated as tumor \cite{pcam, camelyon}.

\paragraph{MHIST} The MHIST dataset consists of 3,152 fixed-size H\&E images ($224\times224$ pixels at 8x magnification) of colorectal polyps. The task is to differentiate between hyperplastic polyp (HP) and sessile serrated adenoma (SSA). Each image is labeled according to the majority vote of seven pathologists. HPs are typically benign, while SSAs are precancerous lesions that can turn into cancer if left untreated and require sooner follow-up examinations \cite{mhist}.

\paragraph{CRC-100K} The dataset contains 107,180 H\&E images ($224\times224$ pixels at 20x magnification) extracted from colorectal cancer scans. The task is to predict from each image the following 9 tissue classes: adipose, background, debris, lymphocytes, mucus, smooth muscle, normal colon mucosa, cancer-associated stroma, and colorectal adenocarcinoma epithelium. We use the unnormalized variant of the training dataset in our experiments (NCT-CRC-HE-100K-NONORM) \cite{crc-100k-1, crc-100k-2}. 

\paragraph{TIL Det.} The pan-cancer tumor-infiltrating lymphocyte (TIL) detection dataset \cite{til-detection-ds, champkit} consists of 304,097 (209,221 training and 56,275 test) H\&E images ($100\times100$ pixels at 0.5 mpp) extracted from scans of FFPE sections comprising 23 different cancer types from TCGA. The task is to predict TIL positive images. A TIL count higher than two indicates a positive image.

\paragraph{MSI CRC} Microsatellite instability binary prediction on H\&E colorectal cancer slides. The training split contains 74,726 image patches and the test split 98,904 image patches from TCGA slides. All image sizes are $224\times224$ pixels at 0.5 mpp \cite{msi-kather2019deep, msi-zenodo, champkit}.

\paragraph{MSI STAD} Microsatellite instability binary prediction on H\&E stomach adenocarcinoma slides. The training split contains 80,456 training and the test split 118,008 image patches from TCGA slides. All image sizes are $224\times224$ pixels at 0.5 mpp \cite{msi-kather2019deep, msi-zenodo, champkit}.

\paragraph{H\&E cell classification - Finetuning}
The experiment in Figure \ref{fig:learning}B is based on the same setup and benchmarking data as the experiment in Figure \ref{fig:tme}B described in Section \ref{sec:methods:eval:TME}, hence performs an 8-class cell classification task. However, for additional adaption of the foundation model encoder next to the linear classification layer on top of the final foundation model embedding, we unfreeze the encoder weights during training. To allow for a fair comparison among all benchmarked foundation models, we sweeped over both learning rate and weight decay in the same fixed range and evaluated the models which performed best on a validation split of the training data. For each model the experiment was repeated 3 times with different random seeding.

\end{document}